\documentclass[prd,preprint,showpacs,showkeys,nofootinbib,floatfix]{revtex4-1}

\usepackage{graphicx}
\usepackage{dcolumn}
\usepackage{bm}
\usepackage{commath}
\usepackage{booktabs}
\usepackage{amsmath,amssymb,amsfonts}
\usepackage{hyperref}
\usepackage{textcomp}

\newcommand{\pom}{{\tt I \! P}}
\newcommand{\beq}{\begin{equation}}
\newcommand{\eeq}{\end{equation}}

\begin{document}

\title{Estimations for the Higgs boson production with QCD and EW corrections in exclusive events at the LHC}

\keywords{Higgs boson production, central exclusive diffractive processes, gap survival probability, next-to-leading order accuracy, electroweak corrections}

\author{M. B. Gay Ducati}

\affiliation{High Energy Physics Phenomenology Group, GFPAE, Instituto de F\'isica, Universidade Federal do Rio Grande do Sul, Caixa Postal 15051, CEP 91501-970, Porto Alegre, RS, Brazil}

\author{G. G. Silveira}

\affiliation{High Energy Physics Phenomenology Group, GFPAE, Instituto de F\'isica, Universidade Federal do Rio Grande do Sul, Caixa Postal 15051, CEP 91501-970, Porto Alegre, RS, Brazil}

\affiliation{Center for Particle Physics and Phenomenology (CP3), Universit\'e catholique de Louvain, B-1348 Louvain-la-Neuve, Belgium}

\begin{abstract}
The Higgs boson production is investigated in proton-proton collisions at next-to-leading-order accuracy in central exclusive diffractive processes at the LHC. The production process by the double Pomeron exchange is analyzed in the diffractive factorization through the Ingelman-Schlein approach, taking into account the parton content of the Pomeron by the diffractive partonic distribution function provided by the H1 Collaboration. Hence, we estimate the production cross section of the Higgs boson as well as its rapidity distribution for distinct energies of the LHC. Also, we include the gap survival probability in our calculation, which is studied in recent works and expected to lie in the range between 1\% and 5\% for the energy regime of 14 TeV. As a result, we found a production cross section of about 0.3--0.8 (1.2--3.7) fb at 7 (14) TeV, being of the same order as predicted by the two-photon and the Balitsky-Fadin-Kuraev-Lipatov Pomeron mechanisms. Therefore, assuming the selection rules of spin-parity properties, the exclusive production is a promising channel for the Higgs boson detection in the LHC.
\end{abstract}

\pacs{12.15.Lk,  12.38.BX,  13.85.Dz,  14.80.Bn}

\maketitle

\section{Introduction}
\label{intro}

The inclusive processes provide the highest cross sections for particle production in hadron colliders, which is expected to be the discovery channel of the Higgs boson at the LHC \cite{Dittmaier:2011ti}. Particularly, the highest contribution for this production mechanism is the gluon fusion vertex $pp \to (gg) \to H + X$ \cite{Georgi:1977gs}, predicting a total cross section of about 15 (50) pb at $\sqrt{s}$ = 7 (14) TeV for a Higgs boson mass of $M_{H}$ = 120 GeV for different methods for the next-to-next-to-leading-order (NNLO) perturbative calculation \cite{Anastasiou:2008tj,*deFlorian:2009hc}. Moreover, there are processes where the final state is the same as the decay products of the Higgs boson, that attenuate the production signal, e.g., the processes $gg \to b\bar{b}$ \cite{Gaemers:1984sj,*Kamal:1995ct}, especially for a Higgs boson mass $M_{H} <$ 135 GeV, and the $gg \to \gamma\gamma$ through the box diagram \cite{Carimalo:1980by,*Dicus:1987fk,*Bern:2001df}. So, these events have to be suppressed in order to enable the Higgs boson detection, and higher-order perturbative calculations are highly necessary.

On the other hand, the central exclusive diffractive (CED) process $pp \to p + [LRG] + H + [LRG] + p$ \cite{HarlandLang:2010ep} is a rich tool to investigate particle production in a cleaner environment and was already studied for the production of $\chi$ mesons \cite{Khoze:2000jm,*HarlandLang:2009qe}, dijets \cite{Martin:1997kv}, diphotons \cite{Khoze:2004ak}, and the Higgs boson \cite{Khoze:1997dr,*Khoze:2000cy}. The main signature of this process is the large rapidity gaps present in the final state, i.e., a region in the rapidity distribution with no hadronic activity, since the interaction by the exchange of Pomerons occurs with no change of the quantum numbers of the particles in collision, with the protons remaining intact in the final state \cite{Bjorken:1992er}. An advantage of this process is the possibility to suppress high background signals by means of the $J_{z}$ selection rule \cite{Khoze:2000jm,Kaidalov:2003fw}, that increases the signal-to-background ($S/B$) ratio for a Higgs boson detection at the LHC. For instance, the exclusive $gg \to b\bar{b}$ process has a cross section of the same order as the exclusive Higgs boson production at the LHC, and then the background process may overcome the production signal \cite{Maciula:2010vc}. Thus, some detectors are going to be set up in the LHC experiments to detect the rapidity gaps, increasing the possibility of observation of such processes \cite{Bonnet:2007pw,*Albrow:2008pn,*Roland:2010ch}.

In addition, secondary interactions occur between the protons during the exclusive reaction that causes a contamination of the final state by other particles, reducing the signal of the CED process. In order to weigh the fraction of processes where the secondary interactions (underlying events) do not fill the rapidity gaps, one has to compute the rapidity gap survival probability (GSP), which is used to obtain the cross section that will be observed experimentally. The studies performed by experimental groups have shown that the use of an overall suppression factor is favored by the phenomenological analysis with the data of diffractive dijet production from the Tevatron experiments \cite{Acosta:2003xi}. There are different groups computing the GSP for the CED Higgs boson production \cite{Khoze:2000vr,*Kaidalov:2001iz,Gotsman:1993vd,*Gotsman:1999xq}, and a probability between 1\% and 5\% is expected for the LHC kinematical regime \cite{Ryskin:2009tk,*Ryskin:2011qe,Gotsman:2011xc}.

There are two different approaches to compute the CED Higgs boson production that depend on the dynamics of the hard Pomeron. The first one consists in the calculation based on nonperturbative \cite{Bialas:1991wj} or perturbative QCD \cite{Khoze:1997dr,*Khoze:2000cy} to compute the scattering amplitude of two-gluon exchange in the $t$-channel. In the latter, the diffractive interaction is set in a way that the gluon fusion vertex is preceded by a soft gluon exchange in order to neutralize the color flow into the bosonic loop, introducing the essence of the Balitsky-Fadin-Kuraev-Lipatov Pomeron: an exchange of two hard gluons in the $t$ channel. Additionally, some phenomenological aspects are also included to consider important features of this process, like off-diagonal unintegrated gluon density, and Sudakov form factors at leading logarithm approximation, that have been recently reexamined \cite{Coughlin:2009tr}.

\begin{figure}
\includegraphics{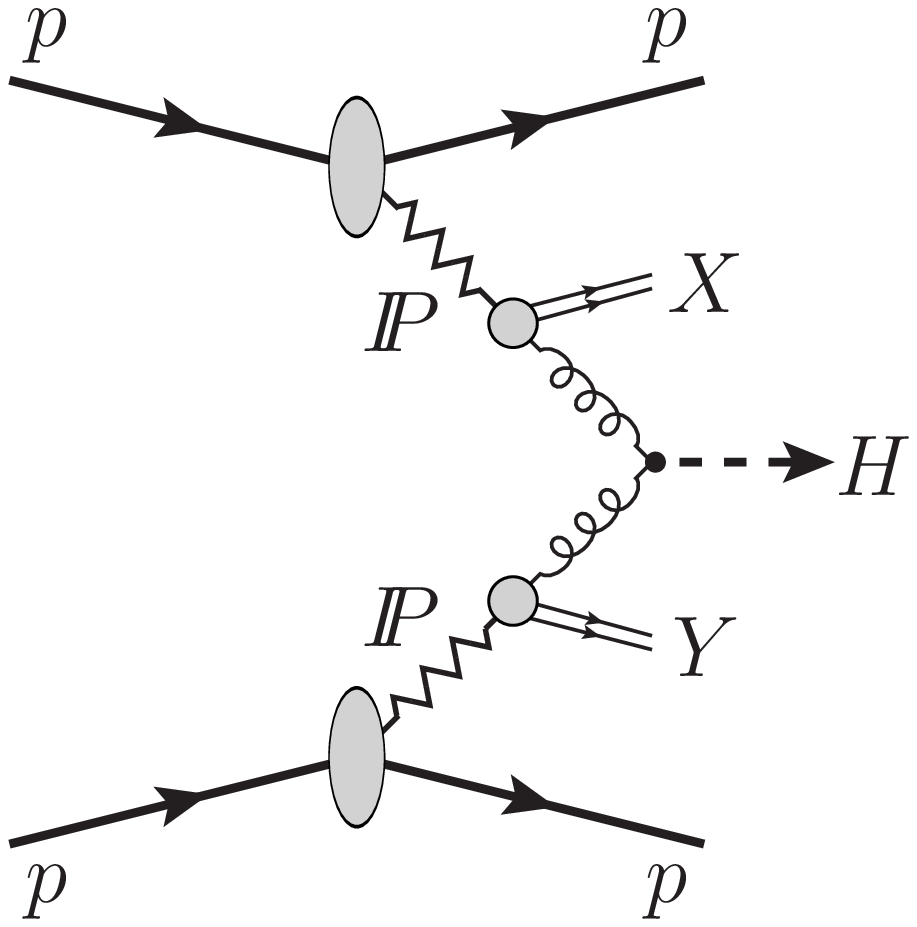}
\caption{\label{diagram}Feynman diagram representing the CED Higgs boson production in the IS approach for proton-proton collisions. The smaller blobs describe the parton content of the Pomeron, which results in the fusion of two gluons and the hadronic states $X$ and $Y$.}
\end{figure}

The second possibility, in which we are mainly interested in this work, was proposed in an analysis of the jet structure in the LEP data \cite{Ingelman:1984ns} and makes it possible to extract the diffractive parton distribution functions of quarks and gluons \cite{Aktas:2006hx,Aktas:2006hy}. Unlike the approach proposed by the Durham group, the Ingelman-Schlein (IS) proposal suggests the emission of Pomerons from the protons and, subsequently, the parton-parton interaction, as presented in Fig.~\ref{diagram}. As the Pomeron is a colorless object, the gluon emission leads to the Pomeron dissociation into hadronic states ($X$ and $Y$ in Fig.~\ref{diagram}), fragmenting into hadrons afterwards.

Therefore, both processes have a similar structure but regard different phenomenological approaches to account for the diffractive interaction to produce the Higgs boson. In this work, we explore the diffractive factorization for the Higgs boson production at next-to-leading order (NLO), estimating the total cross section for the energy regime of the LHC. This study will allow us to confront our predictions with the ones obtained by the Durham group with the use of just a $K$ factor to include all these corrections, since both approaches are evaluated at the same theoretical accuracy\footnote{It is important to note that the Durham group introduces the higher-order contributions to the gluon fusion vertex by a multiplicative factor of 1.5 \cite{Khoze:2000cy}, as verified in Ref.\cite{Spira:1995rr}. It is not an accurate NLO calculation; however, we assume that the NLO contributions are included in that approach, allowing one to perform a direct comparison with our results.}. So, we expect to see wether or not the curves change their behavior with the inclusion of the NLO diagrams. Nevertheless, the addition of the GSP is mandatory in both cases in order to correctly predict the production cross section. As this probability is computed for a specific production mechanism in hadron-hadron collisions, the models for the survival factor employed here can be used in both mechanisms. Thus, this paper is organized as follows: in Sec.~II, we present the inclusive production of the Higgs boson at NLO accuracy. Next, in Sec.~III, the gluon-gluon luminosity is modified to describe the double Pomeron exchange between the protons. Further, in Sec.~IV, the models for the GSP applied in this work are shown, as well as their estimations. In Sec.~V, we discuss the main sources of uncertainties in this approach which affect our predictions. Then, in Sec.~VI, our results are displayed as a function of the Higgs boson mass for different collider energies, as well as the rapidity distributions of the Higgs boson. Finally, in Sec.~VII, we summarize our conclusions.

\section{Inclusive production}
\label{inc}

The inclusive process $pp \to H + X$ at leading-order (LO) accuracy for the gluon fusion vertex corresponds to the fermionic triangular loop \cite{Georgi:1977gs}, for which the quark top has the highest contribution \cite{Djouadi:2005gi}. Following the prescription of hard factorization given in Ref. \cite{Spira:1995rr}, one is able to write the production cross section at LO as
\begin{eqnarray}
\sigma_{LO}(pp \to X + H + Y) = \sigma_{0} \tau_{H} \frac{\dif {\cal{L}}^{gg}}{\dif\tau_{H}},
\label{sigma_LO}
\end{eqnarray}
where $\tau_{H} = M_{H}^{2}/s$ is the Drell-Yan variable of the process, with $s$ the center-of-mass energy squared, and $\sigma_{0}$ is a function of the variable $\tau_{Q} = M_{H}^{2}/4m_{Q}^{2}$, defined as
\begin{eqnarray}
\sigma_{0} = \frac{G_{F}\alpha_{s}^{2}(\mu^{2})}{288\pi\sqrt{2}}\left| \frac{3}{4}\sum_{q} A_{Q}(\tau_{Q}) \right|^{2},
\end{eqnarray}
with $G_{F}$ = 1.16639 $\times$ 10$^{-5}$ the Fermi constant, $\mu$ the renormalization scale, and $A_{Q} = 2[\tau_{Q} + (\tau_{Q} - 1)f(\tau_{Q})]/\tau_{Q}^{2}$. Considering the leading contribution of the top quark for the fermionic loop in the $gg \to H$ vertex, we take the limit $\tau_{Q} \leq 1$ that corresponds to $f(\tau_{Q}) = \arcsin^{2} \sqrt{\tau_{Q}}$ \cite{Spira:1995rr}. The last term in Eq.(\ref{sigma_LO}) is the gluon-gluon luminosity
\begin{eqnarray}
\frac{\dif{\cal{L}}^{gg}}{\dif\tau} = \int_{\tau}^{1} \frac{\dif x}{x} \, g(x,M^{2}) \, g(\tau/x,M^{2}),
\label{lumi}
\end{eqnarray}
where $M$ is the factorization scale, and $g(x,M^{2})$ is the integrated gluon distribution function. This luminosity will play an important role in this work, since it will be used to compute the NLO QCD corrections and will be modified to introduce the Pomeron structure function (PSF) to consider the emission of partons off the Pomeron in the double Pomeron exchange process.

The singular and nonsingular virtual QCD corrections to the $gg \to H$ vertex are expressed through the processes $gg \to H(g)$, $gq \to Hq$, and $q\bar{q} \to Hg$ \cite{Spira:1995rr,Dawson:1990zj}; the NLO contributions to the production cross section in $pp$ collisions can be computed as follows:
\begin{eqnarray}
\sigma_{NLO} (pp \to H + X) = \sigma_{0}\left[ 1 + \frac{{\cal{C}}(\tau_{Q})}{\pi} \, \alpha_{s}(\mu^{2}) \right] \tau_{H} \frac{\dif {\cal{L}}^{gg}}{\dif\tau_{H}} + \Delta\sigma_{gg} + \Delta\sigma_{gq} + \Delta\sigma_{q\bar{q}},
\label{sigma_NLO}
\end{eqnarray}
the singular virtual corrections being expressed in the function ${\cal{C}}(\tau_{Q})$, and the nonsingular ones in the terms $\Delta\sigma_{ij}$, with the renormalization and factorization scales fixed for the strong coupling constant $\alpha_{s}(\mu^{2})$ and gluon distribution functions $g(x,M^{2})$, respectively. In order to add these corrections, the running strong coupling constant has to be input at NLO accuracy. We take into account the exact numerical solution of \cite{Gluck:1998xa}
\begin{eqnarray}
\frac{\dif\alpha_{s}(\mu^{2})}{\dif\ln\mu^{2}} = - \frac{\beta_{0}}{4\pi}\alpha_{s}^{2}(\mu^{2}) - \frac{\beta_{1}}{16\pi^{2}}\alpha_{s}^{3}(\mu^{2}),
\end{eqnarray}
with $\beta_{0} = (11N_{c}-2N_{F})/3$, $\beta_{1} = (102N_{c}-38N_{F})/3$, $N_{c}$ = 3, and $N_{F}$ depending on the quark flavor during the numerical calculation.

The singular virtual corrections correspond to the two-loop corrections, which are expressed by \cite{Spira:1995rr}
\begin{eqnarray}
{\cal{C}}(\tau_{Q}) = \pi^{2} + c(\tau_{Q}) + \left( \frac{11N_{c} - 2N_{F}}{6} \right) \log \frac{\mu^{2}}{M_{H}^{2}},
\end{eqnarray}
where $\pi^{2}$ refers to the infrared part of the cross section for real gluon emissions, and $c(\tau_{Q})$ = 11/2 is solved analytically for $\tau_{Q}~=~M_{H}^{2}/4m_{Q}^{2}$ \cite{Dawson:1990zj,Djouadi:1991tka}. Next, the nonsingular virtual corrections $\Delta\sigma_{ij}$ are obtained from the diagrams of gluon radiation in the $gg$ and $gq$ scattering, and $q\bar{q}$ annihilation. Each of them will be computed through Eq.(\ref{lumi}), modified to include the $q$ and $\bar{q}$ contributions \cite{Spira:1995rr}
\begin{subequations}
\begin{eqnarray}\nonumber
\Delta\sigma_{gg} &=& \int^{1}_{\tau_{H}}\dif\tau\frac{\dif{\cal{L}}^{gg}}{\dif\tau} \sigma_{0} \frac{\alpha_{s}(\mu^{2})}{\pi} \left \{ -\hat{\tau}P_{gg}(\hat{\tau}){\text{log}}\frac{M^{2}}{s} + d_{gg}(\hat{\tau},\tau_{Q}) \right. \\ &+& \left. 12\left [ \left(\frac{{\text{log}}(1-\hat{\tau})}{1-\hat{\tau}}\right )_{+} - \hat{\tau}[2-\hat{\tau}(1-\hat{\tau})]{\text{log}}(1-\hat{\tau}) \right ] \right\}, \\
\Delta\sigma_{gq} &=& \int^{1}_{\tau_H}\dif\tau\sum_{q,\bar{q}}\frac{\dif{\cal{L}}^{gq}}{\dif\tau} \sigma_{0} \frac{\alpha_{s}(\mu^{2})}{\pi}\left \{ d_{gq}(\hat{\tau},\tau_{Q}) + \hat{\tau}P_{gq}(\hat{\tau})\left [ -\frac{1}{2}{\text{log}}\frac{M^{2}}{\hat{s}}+{\text{log}}(1-\hat{\tau})\right]\right\}, \\ 
\Delta\sigma_{q\bar{q}} &=& \int^{1}_{\tau_{H}}\dif\tau \sum_{q}\frac{\dif{\cal{L}}^{q\bar{q}}}{\dif\tau} \sigma_{0} \frac{\alpha_{s}(\mu^{2})}{\pi} d_{q\bar{q}}(\hat{\tau},\tau_{Q}),
\label{Delta_sigma_ij}
\end{eqnarray}
\end{subequations}
as $\hat{\tau} = \tau_{H}/\tau$, and $P_{gg}(\hat{\tau})$ and $P_{gq}(\hat{\tau})$ are the Dokshitzer-Gribov-Lipatov-Altarelli-Parisi splitting functions \cite{Altarelli:1977zs}
\begin{subequations}
\begin{eqnarray}
P_{gg}(\hat{\tau}) &=& 6 \left\{ \left( \frac{1}{1-\hat{\tau}} \right)_{+} + \frac{1}{\hat{\tau}} - 2 + \hat{\tau}(1 - \hat{\tau}) \right\} + \frac{11N_{c}-2N_{F}}{6} \delta(1-\hat{\tau}), \label{ap-eq1} \\
P_{qg}(\hat{\tau}) &=& \frac{4}{3} \frac{1 + (1 - \hat{\tau})^{2}}{\hat{\tau}}.
\label{ap-eq2}
\end{eqnarray}
\end{subequations}
The $F_{+}$ is the plus distribution defined as $F_{+}(\hat{\tau}) = F(\hat{\tau}) - \delta(1 - \hat{\tau}) \int_{0}^{1}d\hat{\tau}^{\prime} F(\hat{\tau}^{\prime})$. It is important to keep in mind the dependence of these individual cross sections on the parton-parton luminosities $\dif{\cal{L}}^{ij}/\dif\tau$, since each of these luminosities will be replaced to introduce the diffractive interaction where the partons are be emitted off the Pomeron.

As we are assuming the top quark contribution in this work, the $d_{ij}(\hat{\tau},\tau_{Q})$ functions can be evaluated numerically \cite{Dawson:1990zj,Djouadi:1991tka}:
\begin{subequations}
\begin{eqnarray}
d_{gg}(\hat{\tau},\tau_{Q}) & = & -\frac{11}{2}(1-\hat{\tau})^{3}, \\
d_{gq}(\hat{\tau},\tau_{Q}) & = & -1+2\hat{\tau}-\frac{\hat{\tau}^{2}}{3}, \\ 
d_{q\bar{q}}(\hat{\tau},\tau_{Q}) & = & \frac{32}{27}(1-\hat{\tau})^{3}.
\label{dfunctions}
\end{eqnarray}
\end{subequations}

Finally, we also account for the electroweak (EW) two-loop corrections to the $gg \to H$ vertex \cite{Actis:2008ts,*Actis:2008ug,*Actis:2008uh}, which increase the production cross section by 5\% in comparison to the NNLO QCD corrections. Then, gathering all corrections to the inclusive Higgs boson production, the total cross section is written as
\begin{eqnarray}
\sigma_{\textrm{tot}} = \sigma_{\textrm{NLO}}(1 + \delta_{\textrm{EW}}).
\end{eqnarray}
Consequently, our predictions reproduce the results obtained in Ref. \cite{Spira:1995rr} for $\sqrt{s}$ = 14 TeV.

However, even with a higher production cross section, the signal from the inclusive production is expected to be strongly attenuated by background processes. So, this is an opportunity to explore the diffractive production as an alternative to detect the Higgs boson at the LHC.

\section{Central Exclusive Diffractive production}
\label{dpe}

Our calculation is based on the IS approach, considering the Pomeron emission off the proton, followed by the interaction of its parton content that produces the Higgs boson, as illustrated by Fig.~\ref{diagram}. This procedure allows one to rewrite Eq.(\ref{lumi}) to replace the gluon densities by the Pomeron flux factor and its partonic distribution function. In this work, we assume a standard Pomeron flux, constrained from the experimental analysis of the diffractive structure function performed by the H1 Collaboration \cite{Aktas:2006hy}. The PSF has been modeled in terms of a light flavor singlet distribution $\Sigma(x)$, i.e., the $u$, $d$, and $s$ quarks and their respective antiquarks. Also, a gluon distribution $g(z)$ is included, with $z$ the longitudinal momentum fraction of the parton in the hard subprocess. The Pomeron trajectory is assumed to be linear, $\alpha_{\pom}(t) = \alpha_{\pom}(0) + \alpha^{\prime}_{\pom}t$, with $\alpha^{\prime}_{\pom}$ and their uncertainties obtained from fits to the data from the H1 detector \cite{Aktas:2006hx}. We choose $x_{\pom} \int^{t_{min}}_{t_{cut}} f_{\pom/p}\dif t = 1$ at $x_{\pom} = 0.003$, where $|t_{min}| \approx m^{2}_{p}x^{2}_{\pom}/(1 - x_{\pom})$ is the minimum kinematically accessible value of $|t|$, $m_{p}$ is the proton mass, and $|t_{cut}|$ = 1 GeV$^{2}$ is the limit of the measurement. The H1 parametrization provides two different inputs for the fit of the partonic structure functions. As our curves show very close results using both fits, we chose the fit A to perform our predictions, applying the cut $x < x_{\pom} \leq 0.05$ in accord with this parametrization.

Considering a Pomeron being emitted from each proton, one can express the luminosity of the CED process as
\begin{eqnarray}
\frac{\dif{\cal{L}}_{\textrm{CED}}^{ij}}{\dif\tau} &=& \int^{1}_{\tau} \frac{\dif x}{x} \int_{x}^{0.05} \frac{\dif x_{\pom}^{\underline{1}}}{x_{\pom}^{\underline{1}}} F_{i/\pom/p}\left(x_{\pom}^{\underline{1}},\frac{x}{x_{\pom}^{\underline{1}}},M^{2}\right) \int_{\tau/x}^{0.05} \frac{\dif x_{\pom}^{\underline{2}}}{x_{\pom}^{\underline{2}}} F_{j/\pom/p}\left(x_{\pom}^{\underline{2}},\frac{\tau}{x_{\pom}^{\underline{2}}x},M^{2}\right),
\label{ced-lumi}
\end{eqnarray}
where $x_{\pom}^{\underline{1}(\underline{2})}$ is the momentum fraction of the Pomeron carried by the parton relative to the hadron 1 (2), and the PSF $F_{i/\pom/p}$ is expressed by
\begin{eqnarray}
F_{i/\pom/ p} = f_{\pom/ p}(x_{\pom})f_{i/\pom}\left (\frac{x}{x_{\pom}},M^{2} \right ),
\label{func_pom}
\end{eqnarray}
$f_{\pom/ p}(x_{\pom})$ being the Pomeron flux, and $f_{i/\pom} (\beta,\mu^{2})$ the parton distribution function into the Pomeron, where $i,j$ stands for $g$, $q$, and $\bar{q}$. Nevertheless, it is necessary to correctly estimate the fraction of events during which the rapidity gaps will not be filled by the particles from underlying events with the use of the survival factor.

In order to include the QCD and EW corrections to the gluon fusion vertex, as obtained in Ref.~\cite{Spira:1995rr,Actis:2008ts,*Actis:2008ug,*Actis:2008uh}, we replace the parton luminosities of all individual cross sections, which keep the use of the $\overline{\mbox{MS}}$ factorization scheme. Moreover, we work in the same accuracy as the H1 2006 parametrization, which evolves the diffractive parton distribution functions with the DGLAP evolution equation at NLO accuracy \cite{Aktas:2006hy}. The Pomeron-to-parton splitting functions, like Eqs.\eqref{ap-eq1} and \eqref{ap-eq2}, are not included in this evolution equation at this accuracy level.

\section{Gap Survival Probability}
\label{gsp}

The computation of the total cross section for the CED Higgs boson production demands the addition of a survival factor, assumed enhanced diagrams for multi-Pomeron interactions in high energies \cite{Ostapchenko:2006nh,*Ostapchenko:2010gt}. The initial studies in this subject started computing the soft Pomeron exchanges through eikonal scattering and later added the enhanced diagrams with the triple Pomeron coupling $G_{3\pom}$, which brings important contributions to the GSP. We adopt the two most discussed models for the GSP in CED Higgs boson production.

Firstly, there is the Kaidalov-Khoze-Martin-Ryskin (KKMR) model \cite{Khoze:2000vr,Kaidalov:2001iz} that assumes enhanced diagrams for Pomeron exchanges, predicting a GSP for the CED Higgs boson production of 2.4\% (1.5\%) at 7 (14) TeV \cite{Ryskin:2009tk,*Ryskin:2011qe}\footnote{Note that the probability of 1.5\% in the KKMR model is seen as a lower limit for the GSP in the CED Higgs boson production at the LHC. This probability becomes higher depending on the Pomeron dynamics under consideration.}. Secondly, there is the Gotsman-Levin-Maor (GLM) model \cite{Gotsman:1993vd,Gotsman:1999xq}, working with the QCD and $N$=4 super Yang-Mills \cite{Gotsman:2009bn,Gotsman:2011xc}. As a result, a probability was found of 6.4\% (4.6\%) at 7 (14) TeV. However, there is a change between the GSP obtained in Refs.\cite{Gotsman:2011xc,Gotsman:2009bn}, since the estimated GSP is increased from 0.15\% to about 3\%. This effect is explained by the inclusion of semienhanced diagrams, that were neglected in previous calculations of the survival factor, and introduces important contributions.

Moreover, there are other approaches that account for the GSP in CED Higgs boson production, indicating similar values as found in the KKMR model \cite{Bartels:2006ea} but also very small probabilities, like 0.44\% \cite{Kozlov:2006cu}. Therefore, we employed the GSP from the KKMR and GLM models, currently in agreement with each other.

\section{Results and comments}
\label{results}

\begin{figure}
\includegraphics{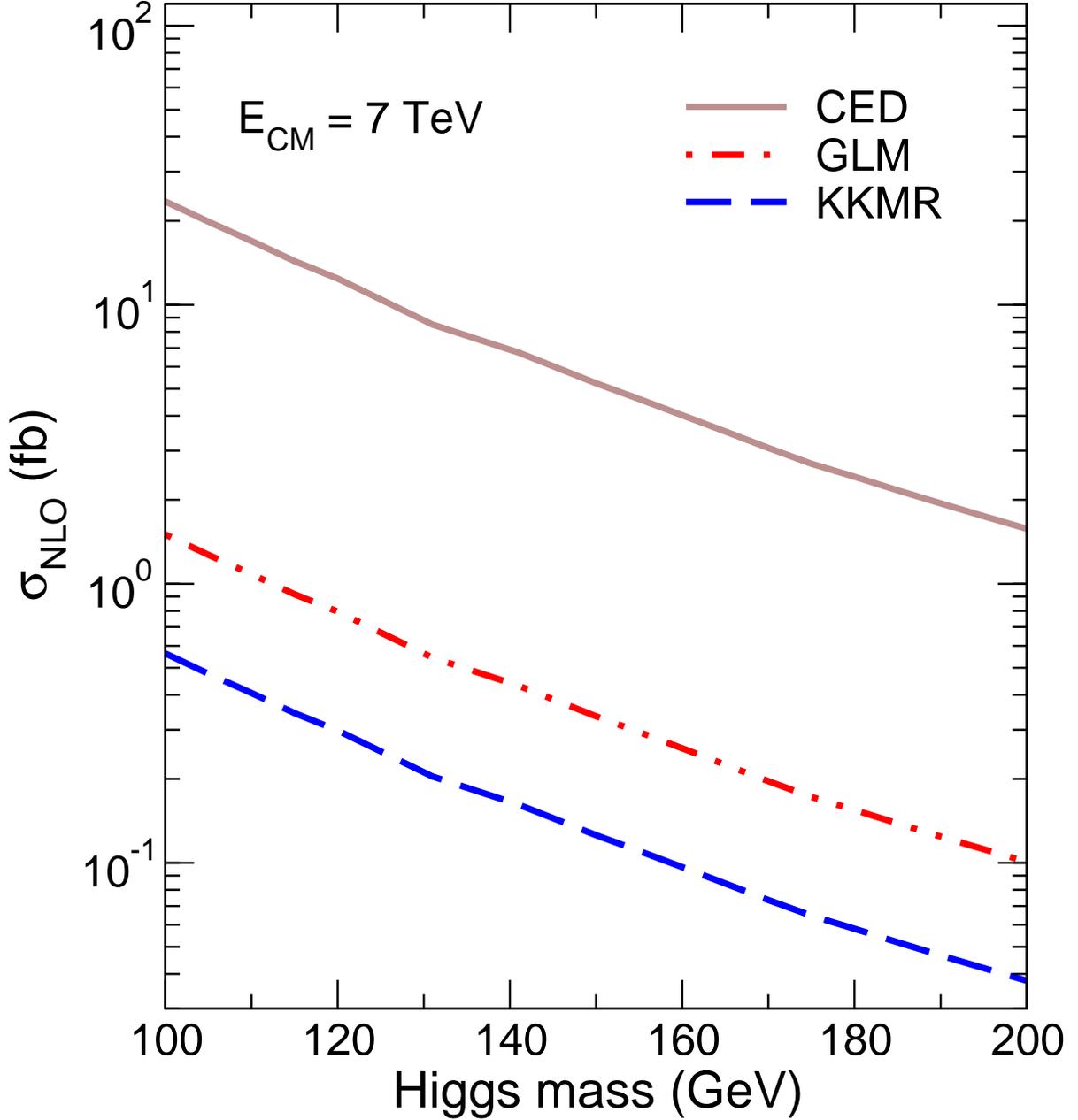}
\caption{\label{CED_mh_7tev}Total cross section in fb in function of the Higgs boson mass for the CED production at $\sqrt{s}$ = 7 TeV with no survival factor (solid line) and taking the GLM model (double-dot-dashed line) and the KKMR model (dashed line) for the GSP.}
\end{figure}

\begin{figure}
\includegraphics{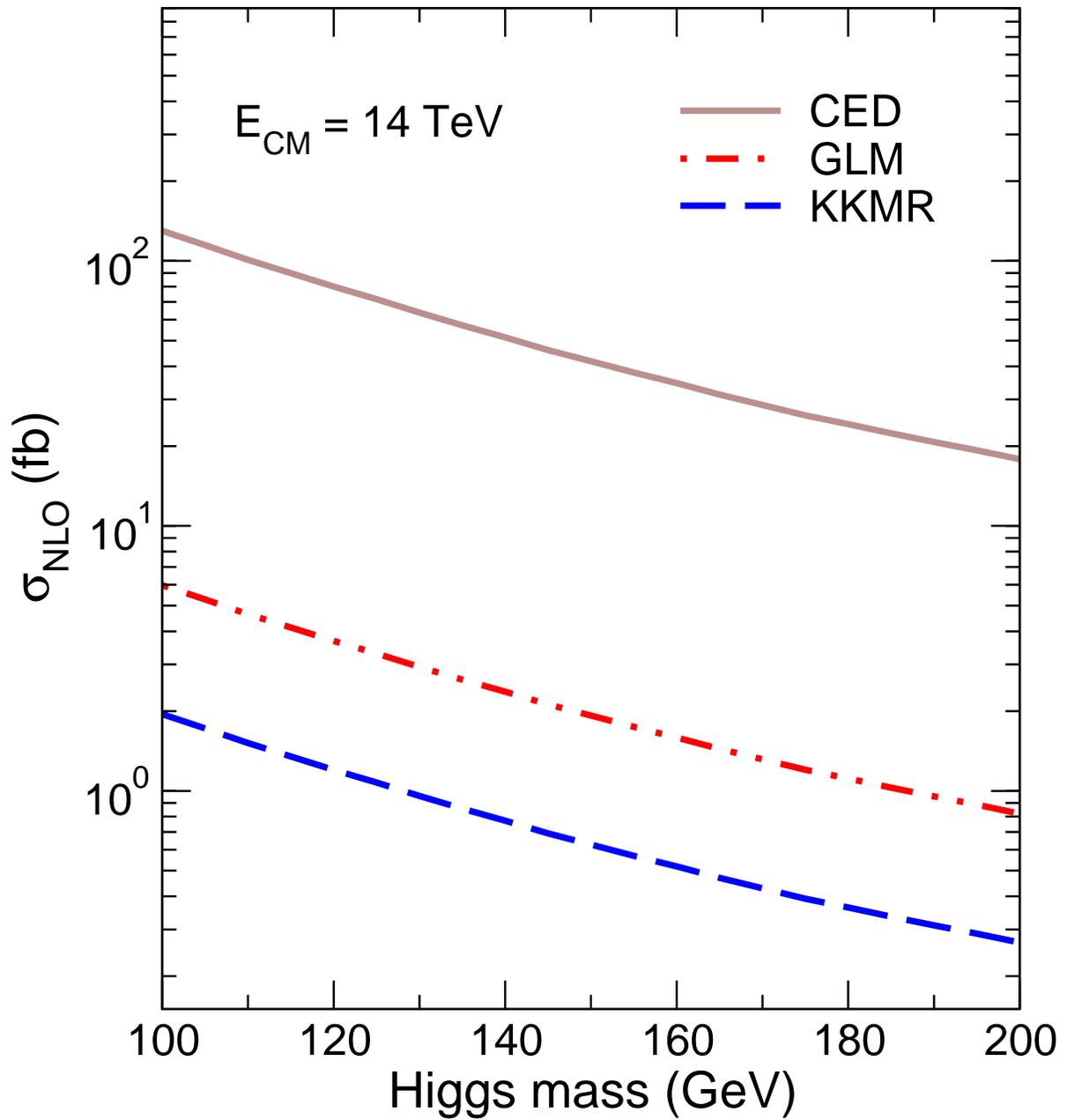}
\caption{\label{CED_mh_14tev}The same as Fig.~\ref{CED_mh_7tev} for $\sqrt{s}$ = 14 TeV.}
\end{figure}

\begin{figure}
\includegraphics{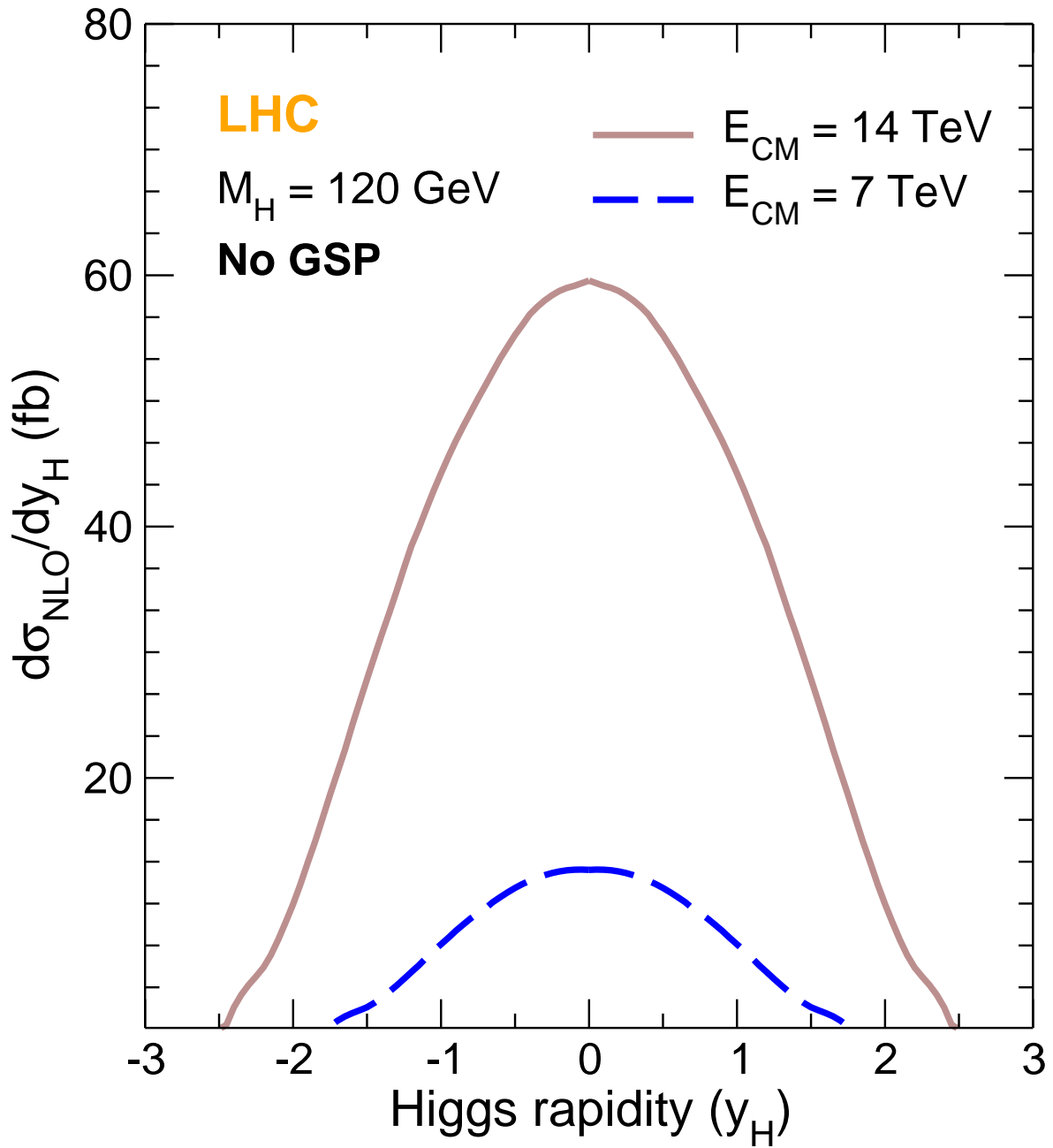}
\caption{\label{CED_yh}Differential cross section in fb in function of the Higgs boson rapidity for $M_{H}$ = 120 GeV in different collider energies. The GSP is not included in this results.}
\end{figure}

Evaluating Eq.(\ref{sigma_NLO}) with the modified luminosity and including electroweak corrections, we have computed the cross section of the CED Higgs boson production for the LHC energies of 7 and 14 TeV. In order to include the suppression of the cross section due to underlying events, we have used the GSP for both kinematical regimes. In Figs.~\ref{CED_mh_7tev} and \ref{CED_mh_14tev}, we show our curves for the total cross section. Moreover, the cross sections using the KKMR and GLM models for the survival factor are also presented. In Fig.~\ref{CED_yh} is displayed the rapidity distribution for a Higgs boson of $M_{H}$ = 120 GeV. As expected, the results lie in the same order as the ones from the Durham group. However, in comparison to our results for different LHC energies, we see a change in the shape of the curves, showing that the NLO diagrams, fully included in this approach, bring distinct contributions to the cross section. This aspect can not be predicted if one uses just a $K$-factor to include such corrections, which introduces further uncertainties to the predictions.

Furthermore, the GSP is the only untested parameter for the exclusive Higgs boson production, causing an uncertainty of more than a factor of 2 if one compares the two GSPs applied to our predictions. As already successfully implemented for the $\chi_{c}$ production in the Tevatron \cite{HarlandLang:2009qe}, the estimations for the survival factor by both models seems to be reliable to be used in the CED Higgs boson production in the LHC kinematical regime.

However, these small cross sections reveal that the detection of the Higgs boson in exclusive processes is going to be a challenge in the LHC, especially because the decay channel $H \to b\bar{b}$ \cite{Carena:2002es,*Hahn:2006my,Djouadi:2005gi}, with a higher branching ratio, is not going to be observed in the ATLAS and CMS experiments \cite{Aad:2009wy,*Ball:2007zza}. Then, other decay channels can be an alternative for this observation, like $H \to \tau^{+}\tau^{-}$, but showing a lower branching ratio if compared to BR($b\bar{b}$). In addition, maybe more feasible, one may look for the Higgs boson decay into a $W^{+}W^{-}$ pair in the range $M_{H} >$ 135 GeV, which has a higher branching ratio but smaller cross section than in the range $M_{H} <$ 135 GeV, as seen in Figs.~\ref{CED_mh_7tev} and \ref{CED_mh_14tev}.

\section{Uncertainties}
\label{concl}

In the use of diffractive factorization, we found different sources of  uncertainties that affect our results. First, the parametrization for  the PSF provided by the H1 Collaboration is constrained with the data  obtained in HERA and does not contribute with significant  uncertainties to the parton-parton luminosities. Next, the survival  factor brings a large uncertainty to our calculations, since it is a  model-dependent approach to account for the soft interactions between  the protons. As presented in Sec.~\ref{gsp}, we apply two distinct  approaches of the GSP for the production cross section, where the  probabilities differ substantially between each other. Next, the  perturbative calculations of the NLO corrections for the gluon fusion  vertex have uncertainties related to the renormalization scale  $\mu_{R}$ in the individual cross sections and also to the  factorization scale $\mu_{F}$, which mainly affects the parton-parton  luminosities.

\begin{figure}
\includegraphics{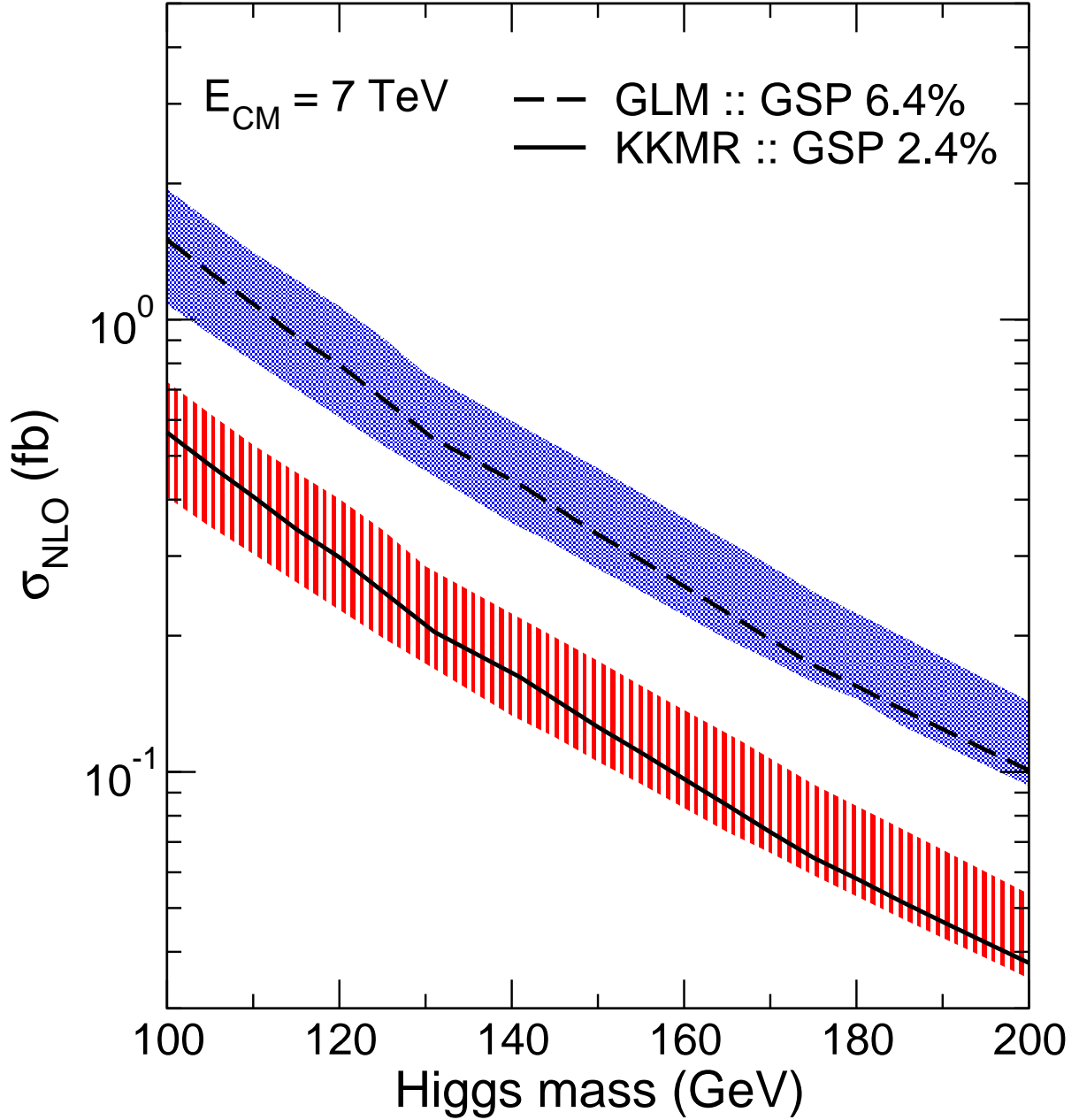}
\caption{\label{fig_uncert}Production cross section in fb in function of the Higgs boson mass considering the variation of the renormalization and factorization scales. The lines represent the predictions for the production cross section with $\mu_{R}$ = $\mu_{F}$ = $M_{H}$ using the survival factors from the GLM model (the dashed line for 3\% and the dot-dashed line for 5\%) and the one from the KKMR model (the solid line for 1.5\%). Both dotted and striped bands represent the results with $\mu_{R}$ = $\mu_{F}$ = \textonehalf$M_{H}$ in the lower limit and $\mu_{R}$ = $\mu_{F}$ = $4M_{H}$ in the upper limit.}
\end{figure}

In order to estimate the total effect of these uncertainties in our  predictions, Fig.~\ref{fig_uncert} shows the variation of the  scales and the GSP with our results for $\sqrt{s}$ = 7 TeV. The upper  curves are the predictions using the GLM model for the survival  factor, with the dashed line representing the probabilities of 6.4\%,  with $\mu_{R}$ = $\mu_{F}$ = $M_{H}$. The dotted band shows the  variation of the production cross section with the scales $\mu_{R}$  and $\mu_{F}$, where the lower limit represents the results with  $\mu_{R}$ = $\mu_{F}$ = \textonehalf$M_{H}$. Moreover, we performed  the same analysis for the production cross section using the survival  factor estimated by the Durham group. The results with a GSP of 2.4\%  are shown in Fig.~\ref{fig_uncert} by the solid line, and the striped  band represents the variation of the scales with the same limits as in  the GLM results.

As a result, the effect of all these parameters is an uncertainty of  5, which is in agreement with the uncertainties found in other  approaches for the CED Higgs boson production  \cite{Dechambre:2011py,*Maciula:2010tv}. The same analysis with $\sqrt{s}$ = 14 TeV leads to a similar effect on the predictions,  showing an uncertainty close to 5. The gap seen in  Fig.~\ref{fig_uncert} between the results for the GLM and KKMR models  is a result of our choice of the GSP from the KKMR model, since the  probability of 2.4\% is taken as the lower bound for the survival  factor in this approach for $\sqrt{s}$ = 7 TeV. In the case of higher values for the survival  factor in the KKMR model, it is expected to see no gap between these  curves.

\section{Conclusions}
\label{concl}

In summary, we have computed the CED Higgs boson production at NLO accuracy for the kinematical regime of the LHC. In this calculation, we have included virtual QCD and electroweak corrections to the CED production of the Higgs boson in the diffractive factorization. As the nature of the underlying events is still an open question in high-energy physics, we have computed our predictions using two distinct models for the GSP in order to explore the possibilities of this production mechanism. Thus, we obtained a production cross section of 0.29 (1.2) fb at $\sqrt{s}$ = 7 (14) TeV for $M_{H}$ = 120 GeV using the KKMR model and 0.79 (3.68) fb at $\sqrt{s}$ = 7 (14) TeV in the GLM model. Our results agree with the ones from the Durham group, obtaining a production cross section of the order of a few femtobarns. Nevertheless, if the GSP is really of the order of 1\%, the two-photon process becomes a promising production mechanism in the LHC with a production cross section of 0.10--0.18~fb \cite{Khoze:2001xm,*d'Enterria:2009er,*Miller:2007pc}, with almost no suppression due to underlying events. Also, the cross section obtained in the photoproduction mechanism lies in the same range \cite{GayDucati:2010xi}, revealing that the scenario for the CED Higgs boson production is very competitive in the LHC energy regime. Therefore, both approaches for the CED Higgs boson production show similar results and may be an opportunity to discover the Higgs boson at the LHC, despite the fact that the observed decay channels will bring experimental difficulties for this detection.

\begin{acknowledgments}
This work was supported by CNPq, Brazil. G.G.S. would like to thank M. M. Machado and M.~V.~T.~Machado for useful comments and the Center for Particle Physics and Phenomenology (CP3) at Universit\'e catholique de Louvain for the hospitality, where part of this work was accomplished.
\end{acknowledgments}

\pagebreak

\end{document}